\documentclass[]{article}

\usepackage{caption}
\usepackage{subcaption}
\usepackage{lmodern}
\usepackage{float}
\usepackage{graphicx}
\usepackage{amssymb,amsmath}

\usepackage[T1]{fontenc}
\usepackage[utf8]{inputenc}

\usepackage{algorithm}
\usepackage{algpseudocode}

\IfFileExists{upquote.sty}{\usepackage{upquote}}{}

\IfFileExists{microtype.sty}{%
\usepackage{microtype}
\UseMicrotypeSet[protrusion]{basicmath} 
}{}

\usepackage[margin=1.5in]{geometry}

\usepackage{hyperref}
\hypersetup{unicode=true,
            pdftitle={Construcción de un Mapa de Vulnerabilidad Sanitaria en Argentina a partir de datos públicos},
            pdfauthor={Antonio Vázquez Brust; Tomás Olego; Germán Rosati},
            pdfborder={0 0 0},
            breaklinks=true}
\usepackage{longtable,booktabs}
\usepackage{graphicx,grffile}
\makeatletter
\def\maxwidth{\ifdim\Gin@nat@width>\linewidth\linewidth\else\Gin@nat@width\fi}
\def\maxheight{\ifdim\Gin@nat@height>\textheight\textheight\else\Gin@nat@height\fi}
\makeatother
\setkeys{Gin}{width=\maxwidth,height=\maxheight,keepaspectratio}

\IfFileExists{parskip.sty}{%
\usepackage{parskip}
}{
\setlength{\parindent}{0pt}
\setlength{\parskip}{6pt plus 2pt minus 1pt}
}

\providecommand{\tightlist}{%
  \setlength{\itemsep}{0pt}\setlength{\parskip}{0pt}}

\usepackage{titling}
\usepackage{authblk}

  \title{Construcción de un Mapa de Vulnerabilidad Sanitaria en Argentina a partir de datos públicos}
    \pretitle{\vspace{\droptitle}\centering\huge}
  \posttitle{\par}

    \author[1]{Antonio Vazquez Brust}
    \author[1]{Tomás Olego}
    \author[1,2]{Germán Rosati}
    \affil[1]{Fundación Bunge y Born}
    \affil[2]{IDAES-Universidad Nacional de San Martín}    

    \preauthor{\centering\large\emph}
  \postauthor{\par}
      \predate{\centering\large\emph}
  \postdate{\par}
    \date{22 de noviembre, 2018}

\begin{document}
\maketitle
\begin{abstract}
	Este documento se propone exponer de forma detallada los criterios de
	procesamiento y las técnicas de análisis utilizadas para la producción
	del \emph{Mapa de Vulnerabilidad Sanitaria} en base al uso de fuentes de datos públicas y abiertas.
	
\end{abstract}

\section{Introducción}\label{introduccion}

Este documento expone los criterios de procesamiento y las técnicas de análisis utilizadas para la producción de un \emph{Mapa de Vulnerabilidad Sanitaria} para la Argentina.

La noción de \textit{vulnerabilidad sanitaria} está relacionada con los llamados determinantes de salud. Dicho de otro modo, existen ciertos factores y variables que se vinculan fuertemente con el estado de salud -en sentido amplio, biológica, psicológica, social- de una persona o población. Estos determinantes pueden ser de tipos diversos y si los mismos se hallan ausentes o en magnitud insuficiente, se produce un estado de vulnerabilidad.

En efecto, se construyó un insumo que busca lograr la identificación de zonas caracterizadas por una alta vulnerabilidad sanitaria, es decir, por el hecho de que no lograr alcanzar un umbral mínimo en el acceso a las prestaciones de salud. En ese sentido, poder articular una definición operativa de \emph{vulnerabilidad sanitaria} que permita abordar y procesar la información disponible es un primer paso importante.

El mapa construido posee un nivel de desagragación elevado: la unidad mínima de análisis es el radio censal. Si bien se da una explicación más detallada en las siguiente secciones, es posible adelantar que un radio censal es la mínima unidad estadística con información pública disponible en que se divide el territorio nacional. Para dar una idea, en zonas urbanas un radio censal puede equivaler a una cuadra. En cambio, en zonas rurales (en las que la densidad poblacional es menor) los radios censales pueden adquirir dimensiones mucho más grandes (inlcuso, llegar a ocupar más de un $km^2$).

El documento se estructura de la siguiente manera:

\begin{itemize}
	\tightlist
	\item
	En la Sección~\ref{definicion-de-vulnerabilidad-sanitaria} se plantean algunas definiciones operativas
	utilizadas para la construcción de la información desplegada en el
	mapa.
	\item
	En las Secciones~\ref{cercania-a-efectores-de-salud} y \ref{nivel-socioeconomico} se listan las fuentes de información utilizadas  se presenta una explicación detallada de los diferentes procedimientos y técnicas aplicadas al procesamiento y análisis de la información de base.
	\item
	La Sección~\ref{generacion-del-indice-de-vulnerabilidad-sanitaria} describe las transformaciones aplicadas sobre las variables con el objetivo de combinarlas en un indicador único.
	\item
	Finalmente discutimos las limitaciones y próximos pasos del trabajo en la Sección~\ref{conclusion}.
\end{itemize}

Cabe destacar que este documento constituye un primer ensayo en esta dirección y, como tal, presenta más preguntas que certerzas, las cuales serán presentadas hacia el final de este documento, en el que se plantean las líneas posibles para el trabajo futuro.

\section{Definición de Vulnerabilidad Sanitaria}
\label{definicion-de-vulnerabilidad-sanitaria}

Existen determinantes de carácter general asociados al estado de salud y que se vinculan al acceso diferencial a servicios y cobertura de salud de diferentes segmentos de la población. El desarrollo de un índice que cuantifica ésta dimensión busca construir un instrumento que permita servir como complemento y contextualizador para el análisis de la prevalencia de ciertas patologías en determinadas zonas. Estimar la \textit{Vulnerabilidad Sanitaria} permitiría identificar y, eventualmente, priorizar "zonas calientes" que presentan bajo acceso a servicios sanitarios. 

La existencia de disparidades en el acceso por parte de la población a los servicios de salud es un fenómeno altamente estudiado y documentado.
En efecto, no todos los estratos de población muestran una probabilidad de acceso y de cobertura médica constante. En líneas generales, los
segmentos de población más pauperizados y/o residentes en zonas aisladas presentan menores niveles de acceso a dichas prestaciones de salud (\cite{beard}, \cite{timyan1993access}).

En la literatura sobre la problemática de salud suele diferenciarse la idea de ``acceso desigual'' de la llamada ``vulnerabilidad'' que remite, en última instancia, al riesgo potencial de desarrollar ciertas enfermedades o a estar expuesto a ciertos factores de riesgo ambientales. 
En este sentido, el estudio de \emph{poblaciones vulnerables} o de \emph{factores de vulnerabilidad} resultan de interés.

Un obstáculo para la cuantificación de ``vulnerabilidad'' emerge de la necesidad de considerar múltiples factores que podrían explicar las desigualdades en el acceso al sistema de salud. En \cite{grabo} se relevan las principales dimensiones e indicadores vinculados
a estas nociones de \emph{vulnerabilidad} y \emph{desigualdades en el acceso a la salud} utilizados en un conjunto de trabajos seleccionados:

\begin{table}[htp]
	\begin{center}
		\begin{tabular}{l r r}
			\hline
			Aspectos de la Vulnerabilidad & Papers & Porcentaje \\
			\hline
			Pobreza & 21 & 91.3 \% \\
			Minoría Racial/Étnica & 18 & 78.3 \%\\
			Enfermedades crónicas (físicas o mentales) & 12 & 52.2 \% \\
			Falta de cobertura de salud & 8 & 34.8 \% \\
			Edad & 6 & 26.0 \% \\
			Encarcelamiento & 3 & 13.0 \% \\
			Condición de migrante & 3 & 13.0 \% \\
			Bajo nivel educativo & 3 & 13.0 \% \\
			Residencia en áreas de baja cobertura & 2 & 8.7 \% \\
			Desempleo & 1 & 4.3 \% \\
			Condición de viudez & 1 & 4.3 \% \\
			Carencia de vivienda & 1 & 4.3 \% \\
			\hline
		\end{tabular}
	\end{center}
	\caption{Indicadores vinculados al concepto de Vulnerabilidad Sanitaria. Tomados de C. Grabovschi, C. Loignon, and M. Fortin: \textit{Mapping the concept of vulnerability related
			to health care disparities: a scoping review}}
	\label{tab:indicators}
\end{table}%

Puede verse en la Tabla ~\ref{tab:indicators} que los principales indicadores utilizados por los estudios analizados en \cite{grabo}
se vinculan a la condición de pobreza, a la pertenencia a minorías étnicas o raciales, a la presencia de enfermedades mentales o físicas de carácter crónico y a la falta de cobertura médica. Cabe destacar que estos indicadores están definidos a nivel de los individuos o personas. Existen también otros factores que determinan el nivel de vulnerabilidad sanitaria que están más vinculados a la dimensión ambiental o del entorno \cite{pruss2016preventing}.

En la aproximación realizada se consideraron varios de los indicadores mencionados. En ese sentido, la definición de \textit{Vulnerabilidad Sanitaria} adoptada en este trabajo se compone de los siguientes factores asociados:

\begin{itemize}
	\item
	\emph{Acceso a prestaciones y servicios de salud por parte del estado:} para esta dimensión se usó como indicador principal la
	cercanía a efectores de salud. Se construyó un dataset que contiene la ubicación (latitud y longitud) de la gran mayoría de los efectores 
	de salud estatales de todo el país. Para ello, se integraron fuentes provenientes del Estado Nacional y de los Estados Provinciales. Se
	calculó el tiempo de caminata desde diversos puntos hasta el Centro de Salud más cercano.
	\item \emph{Nivel Socioeconómico de la Población (NSE):} para construir el NSE, se utilizó información censal. Si bien este punto se detalla 
	más adelante, puede mencionarse que el cálculo del indicador implicó el procesamiento de información censal, correspondiente al CNPyV del año 
	2010, a nivel individuo. Para ello, se seleccionó una serie de variables relevantes -nivel educativo, indicadores de Necesidades Básicas Insatisfechas, etc.- y se combinaron utilizando \textit{variational autoencoders}, un método para la reducción de dimensionalidad
	basado en redes neuronales (ver sección \ref{nivel-socioeconomico}).
\end{itemize}

\section{Metodología de construcción}\label{metodologia-de-construccion}

Las dimensiones anteriormente mencionadas se combinaron para construir un \emph{Mapa de Vulnerabilidad Sanitaria}. El objetivo del mapa de vulnerabilidad es identificar zonas con un déficit potencial en la cobertura sanitaria de la población, es decir, que no  logran superar un umbral mínimo en el acceso a servicios de salud. Teniendo en cuenta este objetivo, se construyó una métrica que permite ordenar y clasificar a las diferentes zonas en función de este déficit potencial.

\subsubsection{Fuentes de información utilizadas}\label{fuentes-de-informacion-utilizadas}

Para la construcción del mapa se recabaron y analizaron las siguientes fuentes:

\begin{itemize}
	\tightlist
	\item
	Datos censales (Censo Nacional de Población, Hogares y Vivienda 2010 \cite{indec_censo1})
	\item
	Polígonos de radios censales
	\item
	Ubicación de Efectores de Salud Pública: hospitales públicos, centros de salud
	y postas sanitarias (en total, 16.564)
	\item
	Ejes de calles (rutas nacionales, provinciales, caminos y trazas
	urbanas) utilizadas para calcular por simulación distancias entre hogares y efectores de salud.
\end{itemize}

\subsubsection{Criterios y técnicas utilizadas para el procesamiento de la
	información}\label{criterios-y-tecnicas-utilizadas-para-el-procesamiento-de-la-informacion}

Como se ha mencionado previamente, si bien se han utilizado datos a nivel de desagregación menor (como por ejemplo, datos individuales del
CNPyV o datos de los efectores de salud), para la construcción del mapa final se procedió a agregar esta información a nivel de radio censal.

En los siguientes apartados detallamos los diferentes procedimientos empleados para la construcción de la información utilizada y las
diferentes técnicas de procesamiento y análisis empleadas.

\section{Cercanía a Hospitales y Centros de Salud}
\label{cercania-a-efectores-de-salud}

\subsection{Construcción y limpieza del dataset de Efectores de
	Salud}\label{construccion-y-limpieza-del-dataset-de-efectores-de-salud}

Para la construcción del indicador \emph{Cercanía a Efectores de Salud} el primer paso fue la construcción de un dataset con registros de ubicación 
de la mayor cantidad posible de efectores de salud en todo el país, localizados con coordenadas de latitud y longitud.

Este dataset fue construido a partir de la integración de diferentes fuentes de datos oficiales:

\begin{itemize}
	\tightlist
	\item
	\emph{Base nacional de Hospitales y Centros de Atención Primaria}: la
	misma fue compilada por el Sistema de Información Sanitaria Argentina
	(SISA - \href{https://sisa.msal.gov.ar/}), obtenido a través del SEDRONAR en el sitio de IDERA
	(\url{http://catalogo.idera.gob.ar}). Este dataset fue utilizado como
	punto de inicio y \emph{base maestra}. El mismo fue enriquecido y
	corregido en base a la información obtenida de fuentes adicionales.
\end{itemize}

Otras fuentes utilizadas:

\begin{itemize}
	\item
	\emph{Efectores de salud del programa SUMAR}: El
	\href{http://programasumar.com.ar/efectores/}{sitio} fue scrapeado\footnote{El término "scraping" o "web scraping" refiere al conjunto de técnicas y tecnologías utilzadas para extraer información de sitios web que no se encuentra debidamente formateada para su extracción}
	para la obtención de los listados de efectores con la dirección de
	cada centro de salud.
	\item
	\emph{Listados de hospitales y centros de atención de salud del
		Programa Nacional de Salud Sexual y Procreación Responsable
		(Ministerio de Salud)}: Se dispone de datos por provincia. Se
	descargaron los datos del
	\href{http://www.msal.gob.ar/saludsexual/centros.php}{sitio} y se
	georreferenciaron.\footnote{El archivo de la provincia de San Juan
		estaba vacío, por lo cual fue eliminado.}\linebreak
\end{itemize}

\textbf{Fuentes provinciales}

Se expone a continuación el listado de fuentes a nivel provincia\footnote{Para el resto de las provincias no se encontraron fuentes alternativas.} exploradas. En general, se trata de Ministerios de Salud provinciales.

\begin{longtable}[]{@{}llll@{}}
	\toprule
	& Fuente (con hipervínculo) & Tarea & Resultado\tabularnewline
	\midrule
	\endhead
	Buenos Aires &
	\href{http://catalogo.datos.gba.gob.ar/dataviews/246454/establecimientos-de-salud-publicos/}{Portal
		datos abiertos} & Descarga & Agrega casos\tabularnewline
	CABA & \href{https://data.buenosaires.gob.ar/dataset/hospitales}{Portal
		datos abiertos} & Descarga & Agrega casos\tabularnewline
	Córdoba &
	\href{https://gobiernoabierto.cordoba.gob.ar/data/datos-abiertos/categoria/salud/mapa-de-los-centros-de-salud/3}{Portal
		datos abiertos} & Descarga & Agrega casos\tabularnewline
	Entre Ríos &
	\href{http://www.entrerios.gov.ar/msalud/?page_id=310}{Ministerio de
		Salud} & Revisión & No agrega casos\tabularnewline
	Formosa &
	\href{https://www.formosa.gob.ar/planacer/hospitalesycentros}{Plan
		Nacer} & Revisión & No agrega casos\tabularnewline
	Mendoza &
	\href{http://www.salud.mendoza.gov.ar/contactos/centros-de-salud/}{Ministerio
		de Salud} & Revisión & No agrega casos\tabularnewline
	Misiones & \href{https://salud.misiones.gob.ar/hospitales/}{Ministerio
		de Salud} & &\tabularnewline
	Neuquén &
	\href{http://www.saludneuquen.gob.ar/programas-y-comites/departamento-de-salud-infantil-y-adolescencia/detencion-precoz-de-hipoacusias/mapa-geografico-con-rutas-rios-lagos-y-sistema-de-salud-neuquen-con-zonas-sanitarias-hospitales-y-areas-programa-enero-2016-sin-fondo/}{Ministerio
		de Salud} & Revisión & Alto esfuerzo\footnote{Datos en formato gráfico.
		No era trivial el esfuerzo requerido para transcribir y georeferenciar
		los datos.}\tabularnewline
	San Luis & \href{http://www.salud.sanluis.gov.ar/mapa/}{Ministerio de
		Salud} & Revisión & No agrega casos\tabularnewline
	Santa Cruz &
	\href{http://saludsantacruz.gob.ar/portal/institucional/hospitales-y-centros-de-salud/}{Ministerio
		de Salud} & Revisión & Pendiente\tabularnewline
	Santa Fe &
	\href{https://salud.santafe.gov.ar/sims/subportal/nomina/nomina.php}{Ministerio
		de Salud} & Scrap & Agrega casos\tabularnewline
	Santiago del Estero &
	\href{http://www.msaludsgo.gov.ar/web2/index.php?cargar=articulo\&id=79}{Ministerio
		de Salud} & Revisión & Pendiente\tabularnewline
	Tierra del Fuego &
	\href{http://www.saludtdf.gob.ar/centros-de-atencion-rio-grande/}{Ministerio
		de Salud - Río Grande} & Revisión & No agrega casos\tabularnewline
	&
	\href{http://www.saludtdf.gob.ar/centros-de-atencion-tolhuin/}{Ministerio
		de Salud - Tolhuin} & Revisión & No agrega casos\tabularnewline
	&
	\href{http://www.saludtdf.gob.ar/http://www.saludtdf.gob.ar/centros-de-atencion-ushuaia/}{Ministerio
		de Salud - Ushuaia} & Revisión & No agrega casos\tabularnewline
	\bottomrule
	\\
	\caption{Listado de fuentes provinciales de efectores de salud}
	\label{tab:cercania}
\end{longtable}

En todos los casos, se aplicó el siguiente procedimiento:

\begin{enumerate}
	\def\labelenumi{\arabic{enumi}.}
	\tightlist
	\item
	Se descargaron archivos consolidados (si los hubiera) o se recolectó el contenido del sitio correspondiente
	\item
	Se georreferenciaron las direcciones de los efectores de salud
	utilizando la API de GoogleMaps.
	\item
	Los registros sin coordenadas fueron descartados.
	\item
	Se ``superpusieron'' los datos resultantes en el mapa junto con el
	dataset inicial.
	\item
	Se retuvieron los puntos que no se superponen -más allá de un buffer
	de 100 metros-.
\end{enumerate}

A título de ejemplo, se presentan los diferentes puntos provenientes de cada una de las fuentes para la Ciudad Autónoma de Buenos
Aires (CABA).

\begin{figure}[h]
	
	{\centering \includegraphics[width=300px]{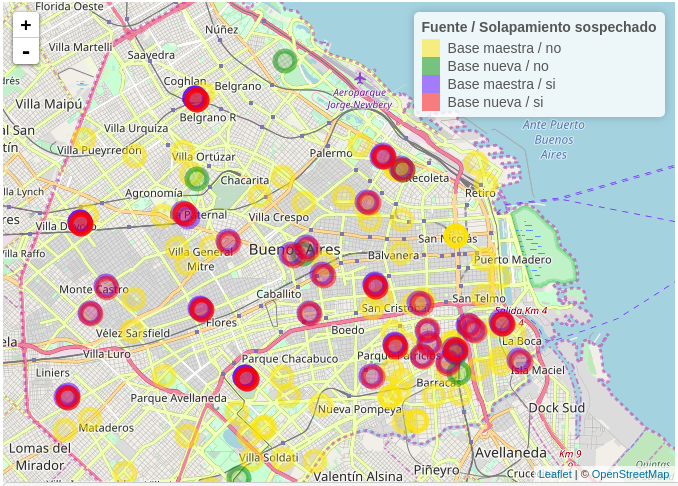} 
		
	}
	
	\caption{Comparación entre la base maestra (SISA) y las diferentes fuentes consultadas para la CABA.}\label{fig:fig2}
\end{figure}

Puede notarse que el dataset de CABA fue notablemente enriquecido: si bien existen bastantes puntos de superposición (rojos y azules), existe
una considerable cantidad de puntos que fueron agregados a la \emph{base maestra} (amarillos).

Al completarse el procedimiento con cada una de las fuentes de datos adicionales adquiridas, el número de efectores de salud pública
georeferenciados aumentó de los 4.419 a 16.564.

\subsection{Clasificación de los efectores de salud según nivel de complejidad}\label{clasificacion-de-los-efectores-de-salud-segun-nivel-de-complejidad}

Con ello se completó la compilación del dataset de efectores de salud en todo el país. Posteriormente, y como última etapa del proceso de limpieza del dataset, los efectores de salud fueron clasificados según su nivel de complejidad. Se buscó el modo de reflejar que la cercanía a un hospital de alta complejidad implica un acceso a prestaciones de salud potencialmente mayor que la cercanía a una posta sanitaria o una ``salita''. El tipo de problemas, las emergencias atendidas y la atención que pueden brindar estos establecimientos difieren notablemente. Si bien existían en los diferentes datasets utilizados clasificaciones en función de la noción de complejidad, los criterios variaban entre fuentes: las clasificaciones no eran homogéneas en los diferentes listados de efectores de salud consultados y consolidados en la base final.

Se realizó un trabajo de revisión con expertos de la Fundación Mundo Sano, que permitió lograr una clasificación que unificara las diferentes denominaciones con los que se describe a los establecimientos de salud, produciendo una clasificación simple según nivel de resolución. El resultado final fue una categorización de los efectores de salud en tres categorías, en orden decreciente de resolución de complejidad en la atención::

\begin{enumerate}
	\def\labelenumi{\arabic{enumi}.}
	\tightlist
	\item
	Hospital
	\item
	Centro de Salud
	\item
	Posta Sanitaria
\end{enumerate}

Tras descartar los efectores de salud pública que no pertenecen a ninguna de las categorías definidas (por ejemplo, geriátricos u oficinas
administrativas), se conservaron 15.903 registros de los 16.654 del total recopilado.

\subsection{Cálculo del tiempo al centro de salud más cercano}\label{calculo-del-tiempo-al-centro-de-salud-mas-cercano}

El paso siguiente en el proceso fue el cálculo del tiempo necesario para llegar al efector de salud más cercano. A partir de este punto, se
comenzó a agregar la información a nivel radio censal.

Era necesario hallar el efector público más cercano a cada radio censal. Este punto representa un problema dado que los radios son polígonos y los efectores, puntos. Una primera opción es utilizar el centroide del radio y calcular la distancia y el tiempo desde esas coordenadas\footnote{La noción de accesibilidad ha sido recientemente utilizada
	como un indicador \emph{proxy} de la distribución desigual de bienes y recursos en
	las sociedad. Un paper reciente\cite{weiss} ha mostrado que existe una correlación positiva y
	relevante entre los tiempos agregados de acceso a aglomerados urbanos de las diferentes
	regiones del mundo y muchos de los indicadores socioeconómicos habitualmente utilizados 
	-niveles de ingreso medios, nivel educativo de la población, tasas de mortalidad infantil,
	etc.-}.

Sin embargo, dado que la forma, los límites y la superficie de los radios censales son muy disímiles a lo largo del país (especialmente, en zonas rurales o poco pobladas) se decidió calcular las distancias y tiempos de la siguiente forma:

\begin{enumerate}
	\def\labelenumi{\arabic{enumi}.}
	\tightlist
	\item
	Dentro de cada radio se seleccionan 5 puntos (pares de coordenadas) al
	azar
	\item
	Se identificó para cada punto el efector de salud más cercano
	\item
	Se calcula para cada punto la distancia/tiempo al efector de salud más
	cercano
	\item
	Se promedian las 5 distancias/tiempos y se obtiene el valor final
\end{enumerate}

Este procedimiento se realizó para cada una de las categorías de efectores de salud -Hospital, Centro de Salud y Posta
Sanitaria. Ahora bien, para el paso 2. se utilizó un algoritmo kNN -\(k\) vecinos cercanos-. Las distancias en metros representadas por diferencias
iguales en latitud y longitud difieren incrementalmente a medida que uno se acerca a los polos y se aleja de la línea de Greenwich. No obstante,
para distancias cortas y suficientemente alejadas de los extremos de la grilla de coordenadas Mercator (tal es el caso en cuestión) el error es
bajo. Por esta razón y para aprovechar la eficiencia de kNN, se aceptó dicho error.

Para la determinación final del tiempo se utilizó Open Source Routing Machine (OSRM), un sistema de ruteo de alta performance que indica las ruta más corta a través de vías públicas entre cualquier par de coordenadas origen-destino (\cite{huber2016calculate}). Para determinar las rutas OSRM utiliza grillas de calles descargadas de OpenStreetMap (\cite{OpenStreetMap}), un repositorio público de información geográfica cuya calidad de datos lo ha establecido como fuente frecuente para estudios de movilidad (\cite{haklay2010good}, \cite{juran2018geospatial})

\subsection{Tiempos a pie}\label{tiempos-a-pie}

El  indicador calculado para la medición del acceso a cobertura de salud fue la distancia a pie hasta el efector de salud más cercano.
Este indicador resulta relevante dado que existe evidencia de que, al menos para cierto tipo de tratamientos médicos, la distancia a pie a un
establecimiento de salud es un buen predictor de las probabilidades de culminación de dicho tratamiento:

\begin{quote}
	Results from this study of Baltimore City clients attending outpatient
	drug addiction treatment programs suggest that having to travel more than 1 mile
	from the treatment center reduces clients' chances of completing
	treatment by almost 50\%, after controlling for the effects of
	demographic variables and type of drug problem. Moreover, living more
	than 4 miles away from treatment decreases the expected length of
	treatment by almost 13 days in comparison to clients traveling less than
	1 mile. \cite{beard} \footnote{
		Resultados de este estudio de pacientes de la ciudad de Baltimore que asisten de forma ambulatoria a
		los programas de tratamiento para la adicción a las drogas sugieren que tener que viajar más de 1 milla
		desde el centro de tratamiento reduce las posibilidades de los clientes de completar
		tratamiento en casi el 50 \%, controlando los efectos de variables demográficas y tipo de problema de drogas. Además, el hecho de vivir a más de 4 millas de distancia del tratamiento disminuye la longitud esperada de tratamiento por casi 13 días en comparación con los pacientes que viajan menos de
		1 milla} 
\end{quote}

En efecto, en el estudio sugiere que una distancia mayor a una milla (aproximadamente 1,6 km.) redunda en un considerable incremento de la probabilidad de no culminación de un tratamiento de rehabilitación. A su vez, distancias mayores a 6.4 km redundan en una baja de la duración media del tratamiento en casi dos semanas.

Se calcularon las distancias a Hospitales, Centros de Salud y Postas Sanitarias

Para un número muy reducido de radios no fue posible determinar un trayecto a pie hasta un efector de salud cercano (10 en total sobre 52.406 radios analizados). Para cada uno de ellos, se realizó inspección visual de imágenes aéreas. En todos los casos se verificó que se trata de radios no urbanizados que abarcan zonas de esteros y bañados, o dedicadas a la agricultura, donde no hay vías peatonales. En esos casos la distancia a pie fue imputada, usando los valores máximos encontrados en el resto de los radios en la misma región administrativa (departamento).

A continuación se presentan mapas a nivel radio censal de los tiempos de llegada a Hospitales, Centros de Salud y Postas Sanitarias para el total del país.

\begin{figure}[H]
	{\centering \includegraphics[width=5.5in]{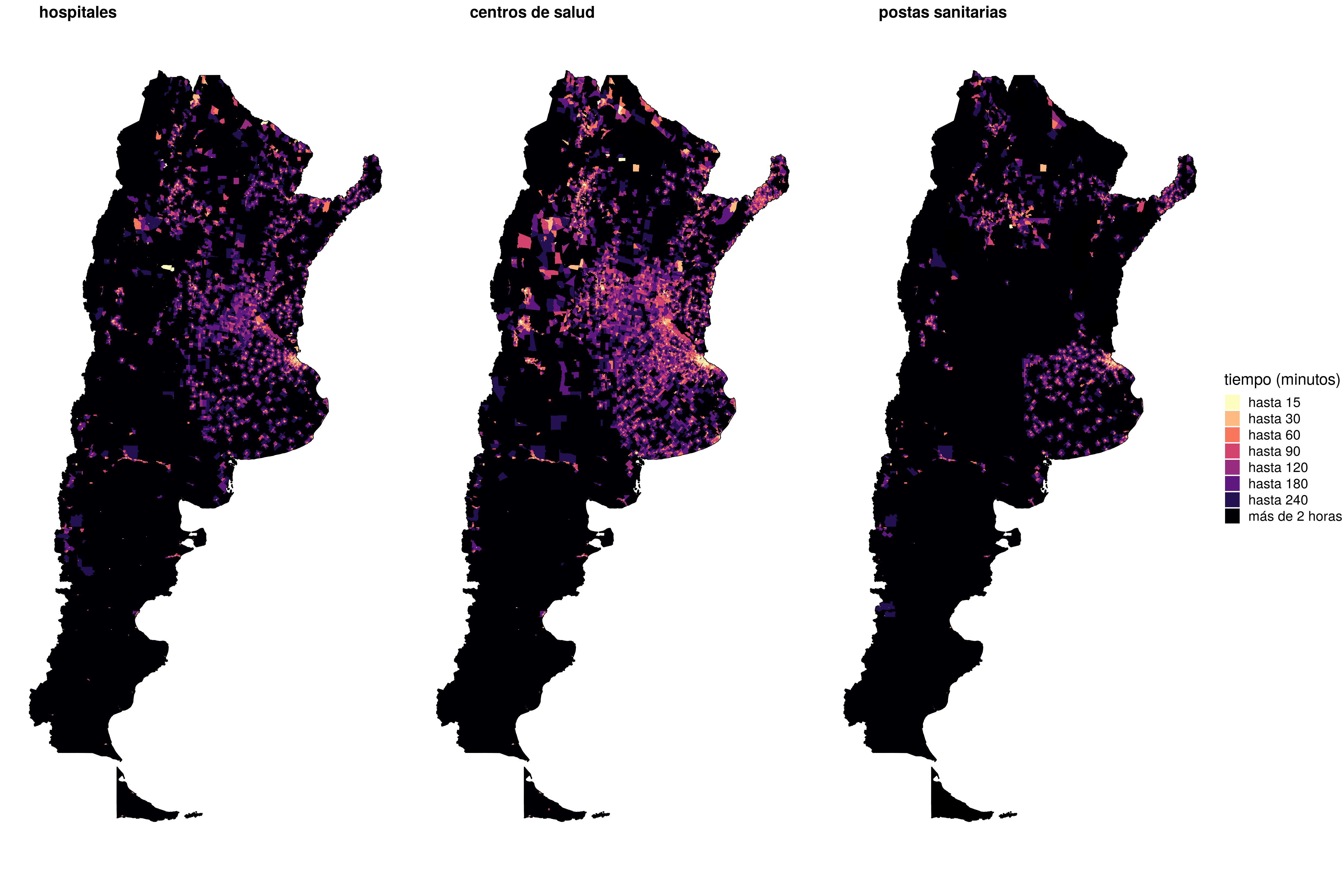}}
	\caption{Tiempos de viaje a pie hasta efectores de salud. Agregados en promedio por radio censal.}
\end{figure}

Como puede verse, los mapas de tiempo de los dos primeros tipos de efectores de salud -Hospitales y Centros de Salud-son notablemente similares. El de Postas Sanitarias resulta más contrastante. En efecto, puede notarse una estructura mucho más segmentada: existe una gran cantidad de regiones con tiempos de viaje notablemente elevado hasta postas sanitarias. De hecho, en una primera inspección visual, pareciera que muchas de estas zonas con altos tiempos de viaje hasta Postas a su vez se caracterizan por valores inversos en tiempos a Hospitales y Centros de Salud.

Esta característica puede deberse a dos factores (no excluyentes entre sí):

\begin{itemize}
	\tightlist
	\item
	\emph{diferentes políticas provinciales en la asignación y asignación
		de postas sanitarias}
	\item
	\emph{errores de cobertura en los datos analizados}, es decir, que
	halla zonas en las que la omisión de postas sanitarias sea muy
	elevada.
\end{itemize}

Quede como una tarea para próximas etapas en esta investigación la determinación del grado de confiabilidad de los datos recopilados en relación a postas sanitarias. 

La distancia final para cada radio $\Delta_{r}$ es  la distancia mediana entre todos los puntos muestreados en cada radio, según ilustra el siguiente mapa:

\begin{equation} 
\Delta_{r} =  MED(\Delta\_{hosp_{1}}, ..., \Delta\_{hosp_{5}}, \Delta\_{centro_{i1}}, ... ,\Delta\_{centro_{i5}}, \Delta\_{posta_{i1}}, ..., \Delta\_{posta_{i5}})
\end{equation}

Es decir incluye todos los efectores-distancias en un radio censal $r$.

Aún con las reservas mencionadas respecto a la fiabilidad de los datos, se incluyó la capa de Postas Sanitarias para el cálculo del indicador, dado que la toma de la distancia mediana minimiza el efecto de potenciales \textit{outliers}. 

\section{Nivel Socioeconómico}\label{nivel-socioeconomico}

\subsection{Datos de entrada}\label{datos-de-entrada}

La segunda dimensión de la \textit{Vulnerabilidad Sanitaria} está constituida por una medida resumen de diversas características socioeconómicas de la población. En efecto, como se observaba en la \ref{tab:indicators} puede verse que las condiciones sociales y económicas de una población, resultan un factor relevante al momento de considerar los riesgos sanitarios relativos. 

Para el cálculo del \emph{Índice de Nivel Socioeconómico} (NSE) se utilizaron datos provenientes del CNPyV del año 2010. Se trabajó con datos a nivel individuo. Dado que el NSE suele ser una variable medida a nivel del hogar, se optó por calcular un índice para cada jefe de hogar en el dataset del Censo.

Para la estimación de los valores del índice se utilizaron las siguientes variables (ver Anexo I), las cuales fueron ordinalizadas:

\begin{longtable}[]{@{}ll@{}}
	\toprule
	Variable & Unidad \tabularnewline
	\midrule
	\endhead
	Condición de propiedad de la vivienda   & Vivienda         \tabularnewline
	Calidad Materiales                      & Vivienda         \tabularnewline
	Calidad de Conexión a Servicios Básicos & Vivienda         \tabularnewline
	Calidad de Construcción                 & Vivienda         \tabularnewline
	Hacinamiento                            & Hogar            \tabularnewline
	Presencia de algún indicador NBI        & Hogar            \tabularnewline
	Nivel Educativo del total del Hogar     & Hogar            \tabularnewline
	Cantidad de Desocupados en el Hogar     & Hogar            \tabularnewline
	Existencia de servicios doméstico       & Hogar            \tabularnewline
	Condición de actividad                  & Individuo (jefe) \tabularnewline
	Nivel educativo                         & Individuo (jefe) \tabularnewline
	\bottomrule 
	\caption{Indicadores utilizados para la construcción del INSE}
\end{longtable}

Para construir el NSE, se utilizó una codificación de termómetro -\textit{thermometer encoding} para las variables ordinales. Sea $N$ la cantidad de casos y $v_1, \ldots, v_ I$ las variables.  Para cada variable $v_i$, existen \(K_i\) categorías. Se crearon las siguientes variables codificadas $x^{(i)}_{k_i}$ para cada variable $v_i$ y para cada categoría $k_i$ donde $ 2 \leq k_i \leq K_i$. En cada caso $j$ con $1 \leq j \leq N$ vale:

\begin{equation}
x^{(i)}_{k_i}(j)=
\begin{cases}
0, & \ si \ v_i(j) < k_i \\
1, & \ si \ v_i(j) \geq k_i
\end{cases} 
\end{equation}

Supongamos que tenemos una variable como el INMAT\footnote{Puede verse una descripción detallada de esta variable en el Anexo 1.}, con cuatro
categorías:

\begin{longtable}[]{@{}llll@{}}
	\toprule
	INMAT original & INMAT\_2 & INMAT\_3 & INMAT\_4\tabularnewline
	\midrule
	\endhead
	1 & 0 & 0 & 0\tabularnewline
	2 & 1 & 0 & 0\tabularnewline
	3 & 1 & 1 & 0\tabularnewline
	4 & 1 & 1 & 1\tabularnewline
	\bottomrule
	\caption{Ejemplo de codificación termómetro}
\end{longtable}

En este ejemplo, INMAT\_1 sería el \emph{baseline} o categoría de referencia, y tomaría un valor constante, por eso es redundante.

\subsection{Construcción del Nivel Socioeconómico}\label{construccion-del-indice}

Para la construcción del índice final se utilizó una técnica de reducción de dimensionalidad llamada \textit{autoencoder}~\cite{goodfellow}. Los \textit{autoencoders} son una arquitectura basada en redes neuronales. En líneas generales, un autoencoder tiene como objetivo encontrar una representación de los datos de input (\emph{encoding}) generalmente con el objetivo de realizar una reducción de la dimensionalidad. En general, los autoencoders funcionan simplemente aprendiendo a replicar los inputs en los outputs. Si bien esto parece un problema trivial introduciendo diversas restricciones a la red puede hacerse esta tarea muy compleja. Por ejemplo, puede limitarse el tamaño de la representación interna o bien agregarse ruido a los inputs. Los autoencoders intentan aprender la función identidad bajo ciertas restricciones \cite{geron}. 

Un autoencoder está compuesto de dos partes:

\begin{itemize}
	\tightlist
	\item
	un encoder (o \emph{recognition network}) que convierte los inputs a
	una representación interna, seguida
	\item
	un decoder (o \emph{generative network}) que reconvierte la
	representación interna a los outputs.
\end{itemize}

Suele tener una arquitectura análoga a un Multi-Layer Perceptron, con la única excepción de que el número de neuronas en la capa de output debe ser
igual a la de input.

El modelo entrenado posee como capa final una función logística y realiza \textit{dropout} (de 0.5) en sus capas intermedias para regularizar y
lograr coeficientes con buena capacidad de generalización.

\begin{figure}[h]
	{\centering \includegraphics[width=350px]{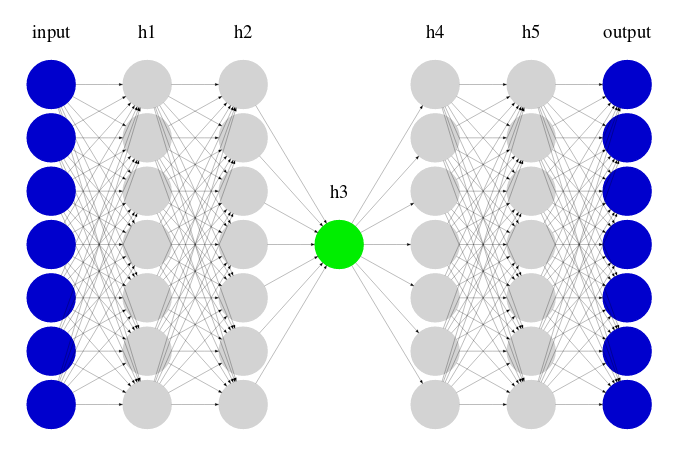} 
	}
	\caption{Esquema del autoencoder utilizado}\label{fig:fig6}
\end{figure}

El modelo que se estimó es el siguiente:

\begin{flalign*}
r_i & \sim \mbox{Bernoulli}(p=0.5) \\
x &= input \\
h_{1} &= \tanh (W_{1} (x * r_1)    + b_{1}) \\
h_{2} &= \tanh (W_{2} (h_{1}* r_2) + b_{2}) \\
h_{3} &= \tanh (W_{3} (h_{2}* r_3) + b_{3}) \\
h_{4} &= \tanh (W_{4} (h_{3}* r_4) + b_{4}) \\
h_{5} &= \tanh (W_{5} (h_{4}* r_5) + b_{5}) \\
\hat{x} &= \sigma (W_{6} h_{5}    + b_{6}) \\
\end{flalign*}
donde $W_i$ es la matriz de coeficientes de la capa $i$, 
$b_i$ es la media de la capa $h$.

De esta forma, la log-probabilidad para cada caso se puede escribir como:
$$
\Pi(\hat{x}, x) =  \sum_{i = 1}^{I} \sum_{k_i = 2}^{K_i}
w^{(i)}_{k_i} \bigg( x^{(i)}_{k_i}  \log \left(\hat{x}^{(i)}_{k_i} \right) + \left(1-x^{(i)}_{k_i} \right) \log \left(1-\hat{x}^{(i)}_{k_i} \right) \bigg)
$$
donde \(w^{(i)}_{k_i}\) se define para cada variable $v_i$ con sus $K_i$ categorías
como:
\begin{flalign*}
w^{(i)}_{k_i} &= \frac{1}{K_i} \cdot \left( \frac{ \sum_{\ell} K_{\ell} }{I} \right)
\end{flalign*}

La función de perdida (\textit{loss function})se definió mediante una verosimilitud pesada:
\begin{flalign*}
L(\hat{x}, x) = \mbox{argmin}_{W, b} & \sum_{j=1}^{N} \Pi(\hat{x}(j), x(j) )
\end{flalign*}

El modelo fue entrenado con ADAM~\cite{geron} definiendo como batchs de datos remuestreados con repetición sobre la distribución empírica de los casos
para favorecer la convergencia.

Como se trabajó con la población completa (ya que el conjunto de datos es el mismo censo), el objetivo del modelo era la generación de una
medida descriptiva. Es por ello que se tomó como criterio de evaluación del modelo la capacidad de explicar a la misma población utilizando como
métrica el promedio ponderado de la probabilidad de cada variable-categoría:

\begin{flalign*}
& Error(\hat{x}, x) = \frac{1}{N} \sum_{j = 1}^{N}  \sum_{i = 1}^{I} \sum_{k_i = 2}^{K_i} 
\frac{1}{\sum_\ell K_\ell}
e^{ x^{(i)}_{k_i}  \log \left(\hat{x}^{(i)}_{k_i} \right) + \left(1-x^{(i)}_{k_i} \right) \log \left(1-\hat{x}^{(i)}_{k_i} \right) } 
\end{flalign*}

El modelo final conserva un 85\% del total de la información de input.

De esta forma, a partir de $h_{3}$ cada jefe de hogar, y por ende, cada hogar, queda clasificado con un valor resultado del autoencoder el cual llamaremos $s_i$, NSE.

A partir del NSE se generó una medida agregada para cada radio tomando como base el nivel socioeconómico que posee cada jefe de hogar. Siendo así, se define para cada jefe de hogar $i$ con un índice de nivel socioecnómico $s_i$ que habita en el radio censal $r$ con una población de $n_r$ jefes de hogar la variable $eta$ como

\begin{equation}
\eta_{r} = \frac{1}{4} Q_{.25}(\mathbf{s}_{r_{1,...,n_r}}) + \frac{1}{2} Q_{.5}(\mathbf{s}_{r_{1,..,n_r}}) + \frac{1}{4} Q_{.75}(\mathbf{s}_{r_{1,..,n_r}})
\end{equation}

donde $Q_{p}$ es el correspondiente $p$ cuantil y $\eta_{r}$ el resultado de aplicar la medida resumen conocida como "Tukey Trimean" la cual logra un compromiso entre robustez y eficiencia comparada a la mediana.

Como puede verse en los density plots previos, el índice construido parece captar en buena medida las disparidades provinciales:

\begin{itemize}
	\tightlist
	\item
	la CABA presenta una distribución claramente sesgada hacia la derecha (valores más altos del INSE)
	\item
	provincias como Chaco, Formosa, Jujuy, Salta presentan distribuciones sesgadas hacia la izquierda.
\end{itemize}

\begin{figure}[h]
	{\centering \includegraphics[width=400px]{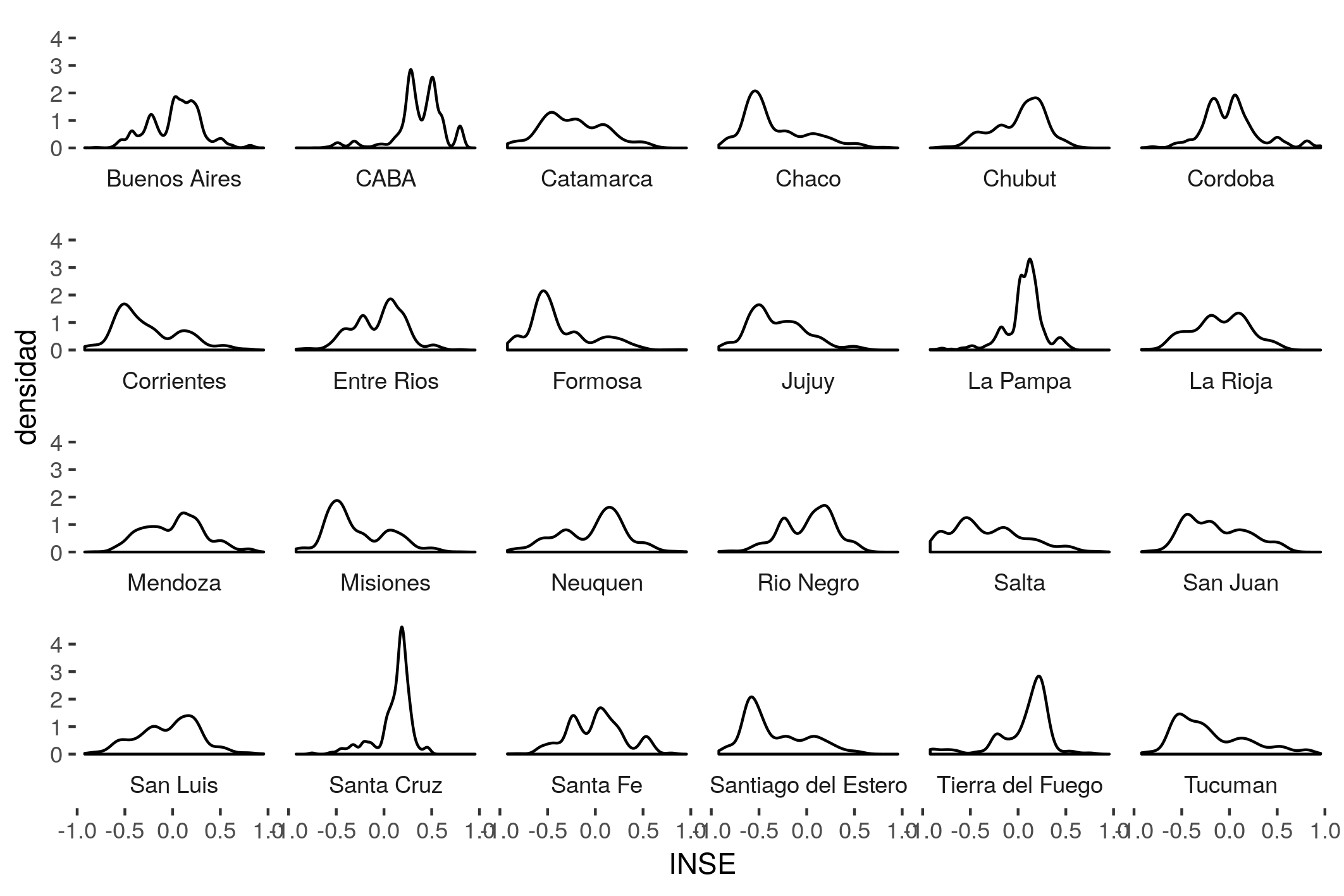} 
	}
	\caption{Density plot del INSE por provincia}\label{fig:dens_plot_INSE_provs}
\end{figure}

\section{Cálculo del Indice de Vulnerabilidad Sanitaria}
\label{generacion-del-indice-de-vulnerabilidad-sanitaria}

\subsection{Dimensiones consideradas}
La construcción del indice de Vulnerabilidad Sanitaria contempló las siguientes dimensiones:

\begin{itemize}
	\item Nivel Socioeconómico
	\item Acceso a Centros de Salud
\end{itemize}

La primera de ellas, como hemos mencionado, puede ser considerada un determinante de carácter general de la situación sanitaria de la población y se refiere al acceso a bienes sociales como la educación y a las condiciones de habitabilidad. Si bien estos bienes se encuentran asociados a trayectorias individuales y sociales, también mantienen una relación con el lugar de residencia. En otras palabras el concepto de nivel socioeconómico está ligado al problema de la distribución y por tanto es una variable intensiva.

Para construir el índice de Vulnerabilidad Sanitaria final es necesario combinar ambas variables a nivel radio censal. La estrategia adoptada para generar un índice compuesto es la de componentes principales, es decir que se busca la combinación lineal de las variables que pueda explicar la máxima varianza.

Ahora bien, cuando la distribución de las variables presentan distribuciones atípicas, es decir multimodales, asimétricas y/o con colas pesadas, la interpretación de los componentes principales resulta dificultosa ya que el método es sensible a la escala de las variables ~\cite{jolliffe2006principal}. Una posible solución es transformar las variables utilizando los rangos de los datos ~\cite{BaxterRankPCA} ~\cite{compositionalRank}

Por ello, siguiendo a ~\cite{solomon}, se buscó estandarizar cada una de las $X_{j}$ variables mediante la transformación rankit:

\begin{equation}
rankit(X_{i,j}) = \frac{r_j(x_{i,j}) - 0.5}{n}
\end{equation}

Para cada observación $i$ de cada variable $X_{j}$ se calcula el rango $r_j(X_{i,j})$ -que varía entre 1 y n-, se le resta 0.5 y se divide por el total de registros $n$. 

De esta manera, se remueven tanto las diferencias de unidades que pudieran existir entre variables, como también se logra invarianza frente a cambios de escala, desplazamientos y transformaciones monótonas. 

\begin{figure}[H]
	{\centering \includegraphics[width=440px, height=440px]{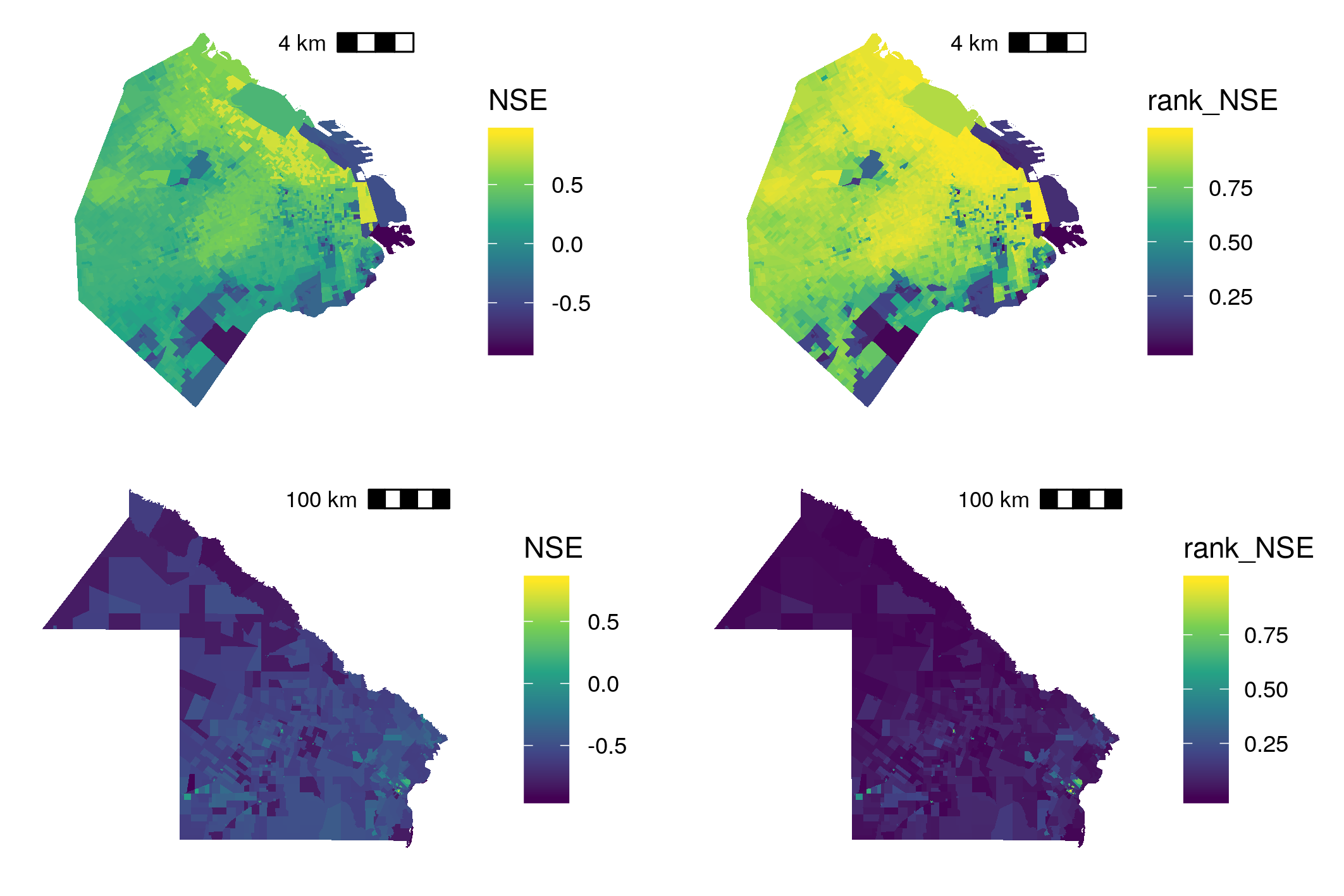} 
	}
	\caption{INSE en escala natural y en escala rankit. CABA y Chaco}\label{fig:nse_chaco_caba}
\end{figure}

\begin{figure}[H]
	{\centering \includegraphics[width=440px, height=440px]{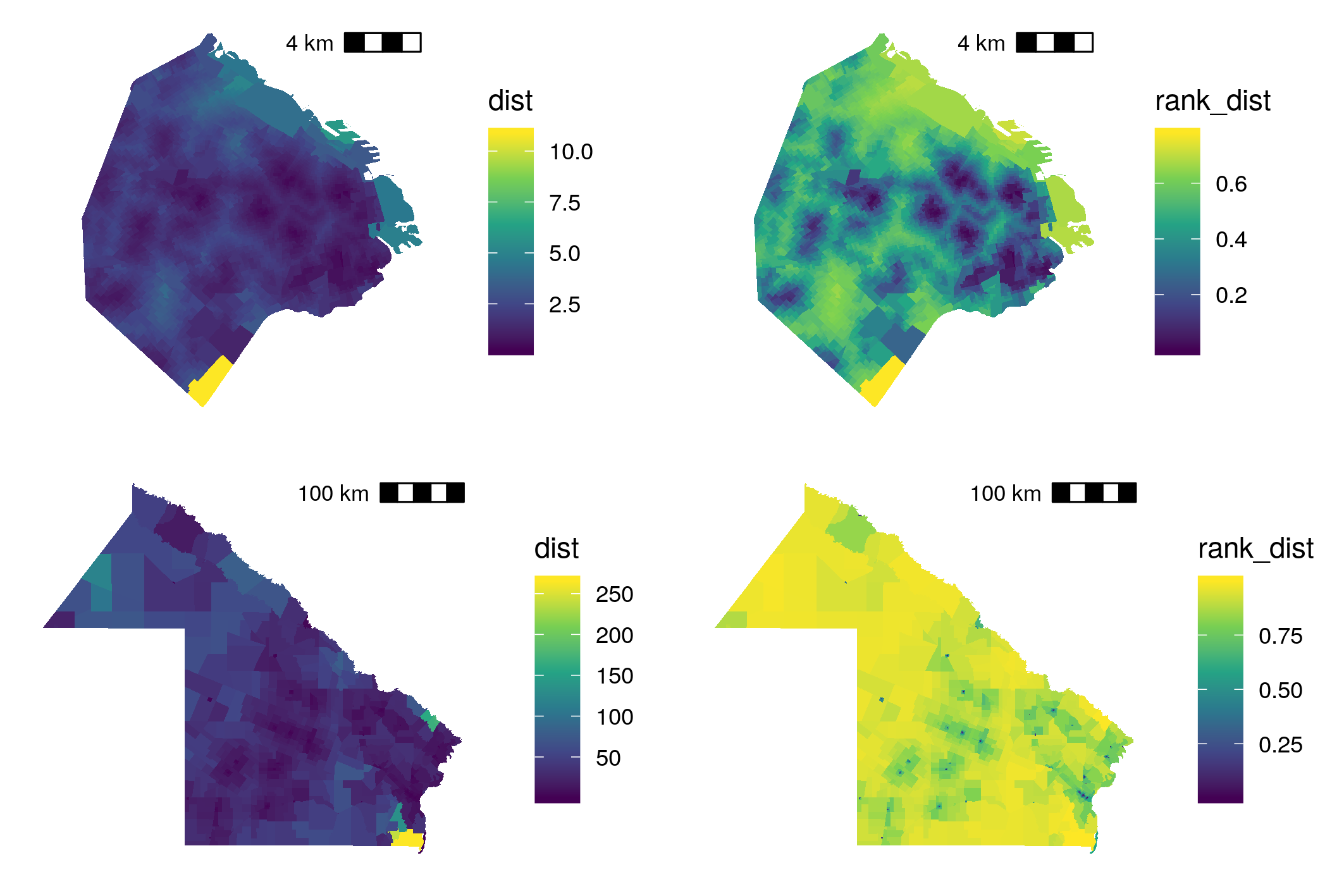} 
	}
	\caption{Distancias a Centros de Salud en escala natural y en escala rankit. CABA y Chaco}\label{fig:dist_chaco_caba}
\end{figure}

Puede notarse en los plots anteriores el efecto que la transformación $rankit$ tiene en las dos variables utilizadas: Nivel Socioeconómico -$\eta_{r}$- y Acceso a Centros de Salud -$\Delta_{r}$-. 

Es fácil notar que la transformación rankit no es otra cosa que un estimador no paramétrico de la función distribución acumulada. Luego, y siguiendo a Egger et all.~\cite{EggerPCA} y a Han y Liu ~\cite{LiuPCA}, se procedió a realizar un análisis de componentes principales semiparamétrico:

\begin{algorithm}
	\caption{Componentes Principales Semiparamétrico}\label{SPCA}
	\begin{algorithmic}[1]
		\Procedure{S-PCA}{X}
		\For{\texttt{j en variables}}
		\For{\texttt{i en casos}}
		\State $z_{i,j}=\Phi^{-1}\big(\frac{r_j(x_{i,j}) - 0.5}{n})\big)$ \Comment{$\Phi^{-1}$ es la inversa de la f.d.a. gaussiana}
		\EndFor
		\EndFor
		\State \textbf{calcular} $\Sigma $ \Comment{Matriz de covarianza}
		\State \textbf{encontrar} $U$, $S$ tal que $\Sigma = \frac{1}{n} US^2U^t$ \Comment{SVD}
		\State \textbf{retornar} $ZU^t$, $S$ \Comment{Coordenadas y autovalores}
		\EndProcedure
	\end{algorithmic}
\end{algorithm}

Para realizar la combinación de ambas variables se calculó la correlación de Spearman sobre los rankits ~\cite{LiuPCA} y se descompuso la matriz de correlación no paramétrica. Se encontró que el principal autovector absorbía el 72\% de la variabilidad de los rankits y que estaba orientado a la dirección inversa de crecimiento mutuo. Este primer autovector fue considerado, entonces, la combinación óptima entre ambas variables. De esta manera, y mediante la aplicación sucesiva de transformaciones no paramétricas, se generó un índice tolerante a contaminación y sobre todo independiente a las características de medición de cada variable de entrada.

El objetivo es obtener un índice entre 0 y 1 cuya distribución sea homogenea para todas las unidades de análisis. Se aplicó nuevamente una transformación del tipo de función de distribución acumulada estimada mediante logsplines usando AIC como criterio de regularización sobre la principal dirección, quedando conformado de esta forma el índice $VS_{r}$ para cada uno de los radios:

A diferencia de rankit, al modelar la distribución con logspline se respetan las distancias relativas.

\begin{figure}[H]
	
	{\centering \includegraphics[width=400px, height=400px]{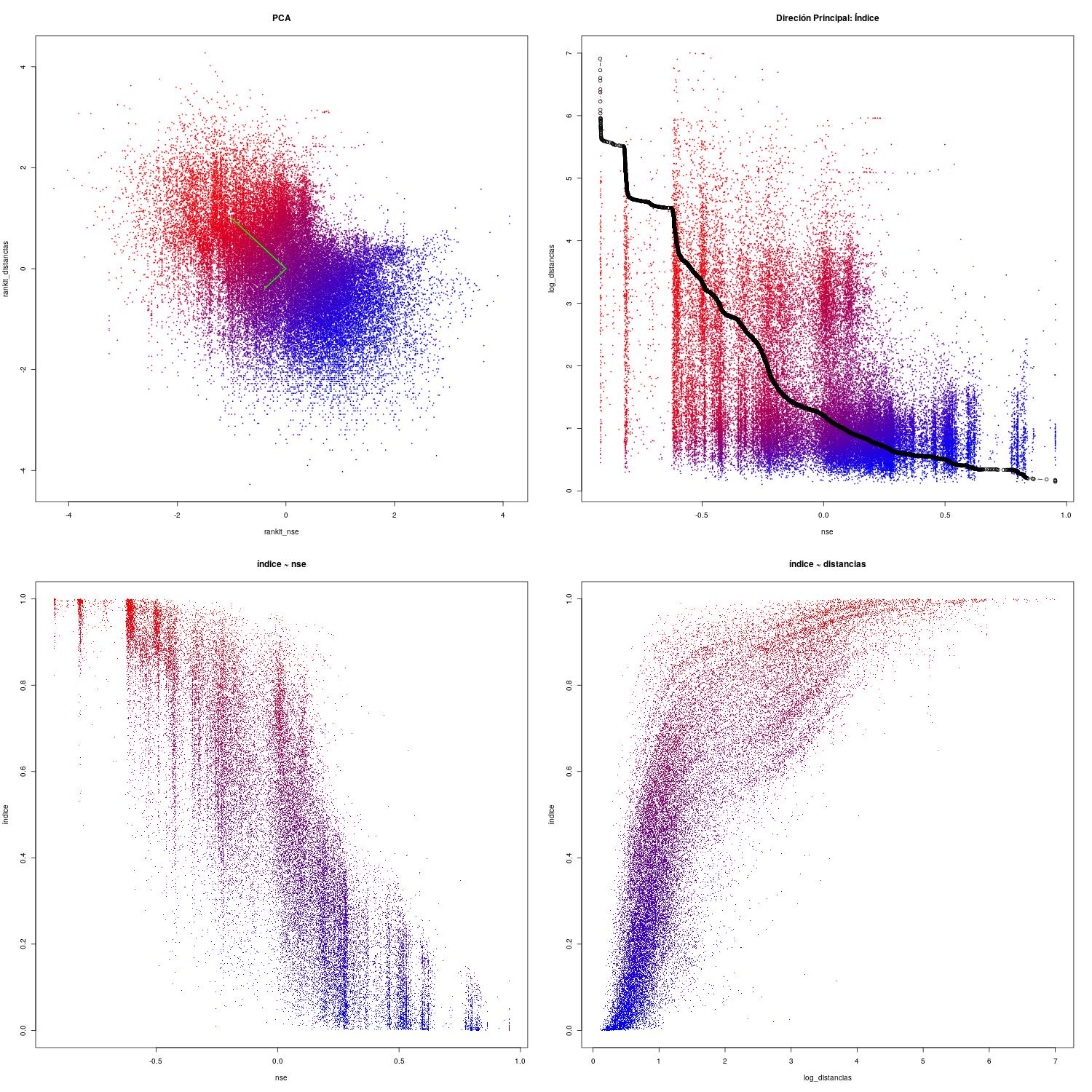} 
	}
	\caption{Índice de Vulnerabilidad Sanitaria}\label{fig:index_vs}
\end{figure}

En el primer gráfico de la figura nueve se observa el resultado del S-PCA en el espacio de las $Z$, y los colores representan el valor del índice resultante. En el segundo gráfico se observa el resultado en el espacio de las $X$. Finalmente en los dos gráficos inferiores se observa el índice en función de INSE y distancias respectivamente.

\begin{figure}[H]
	
	{\centering \includegraphics[]{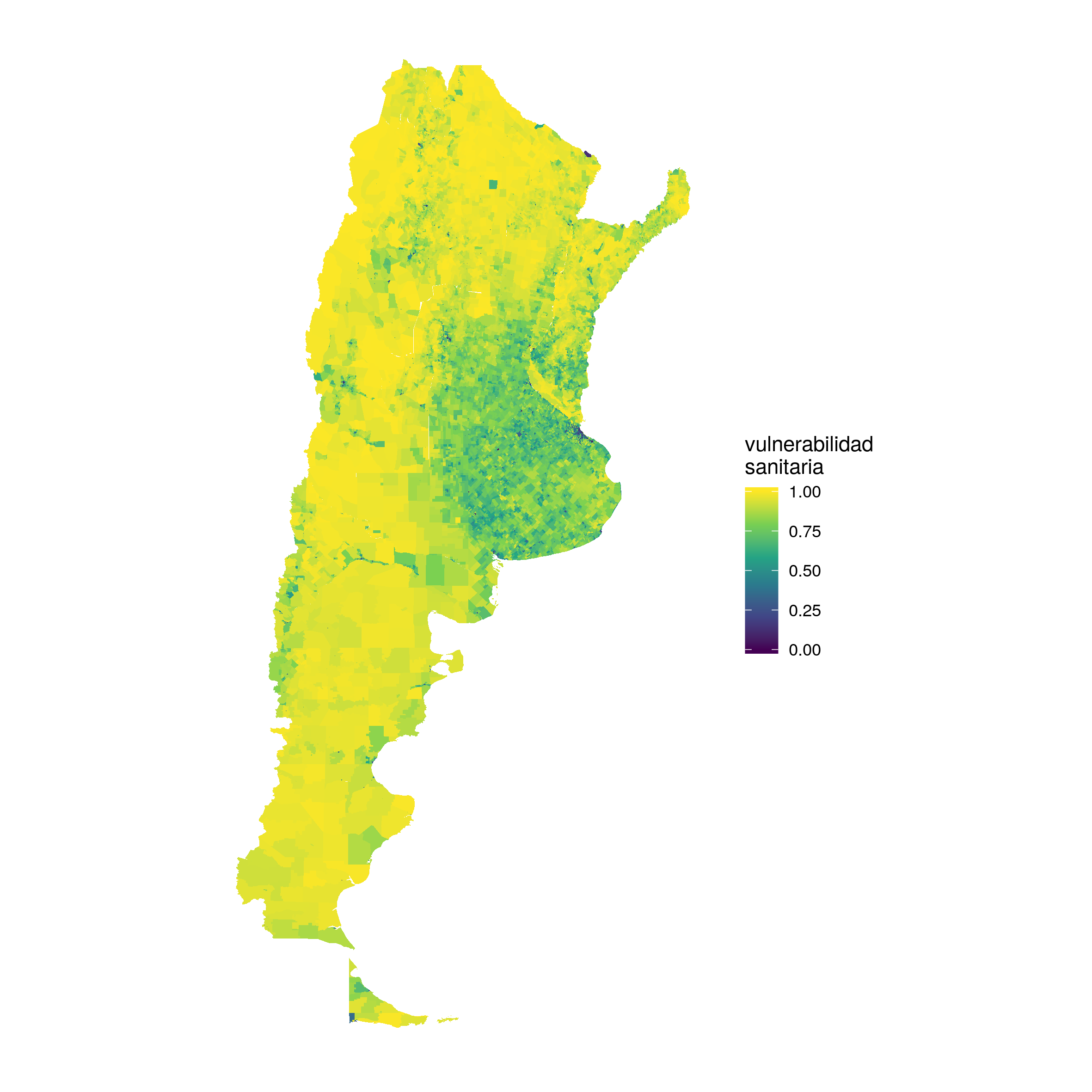} 
	}
	\caption{Índice de Vulnerabilidad Sanitaria a nivel radio censal (Argentina).}\label{fig:indice_vs}
\end{figure}

Ahora bien, esta visualización general debe ser tomada con cuidado debido a la escala (total del país) con la que se está trabajando. En efecto, el mapa anterior parece presentar una situación bastante esperable, dado que es posible detectar dos grandes zonas: 

\begin{itemize}
	\item Región ``central": en esencia Buenos Aires, CABA y los grandes aglomerados urbanos de cada provincia, caracterizada por mayores valores de vulnerabilidad sanitaria,
	\item el resto del país, principalmente las zonas menos densamente pobladas, con valores críticos.
\end{itemize}

Ahora bien, al agregar la información a un nivel superior (fracción censal\footnote{Una fracción censal es la subidivisión inmediatamente anterior al radio censal. Es decir, una fracción está compuesta por un conjunto de radios censales cercanos\cite{indec_censo2}}), el panorama es ligeramente distinto:

\begin{figure}[H]	
	{\centering \includegraphics[]{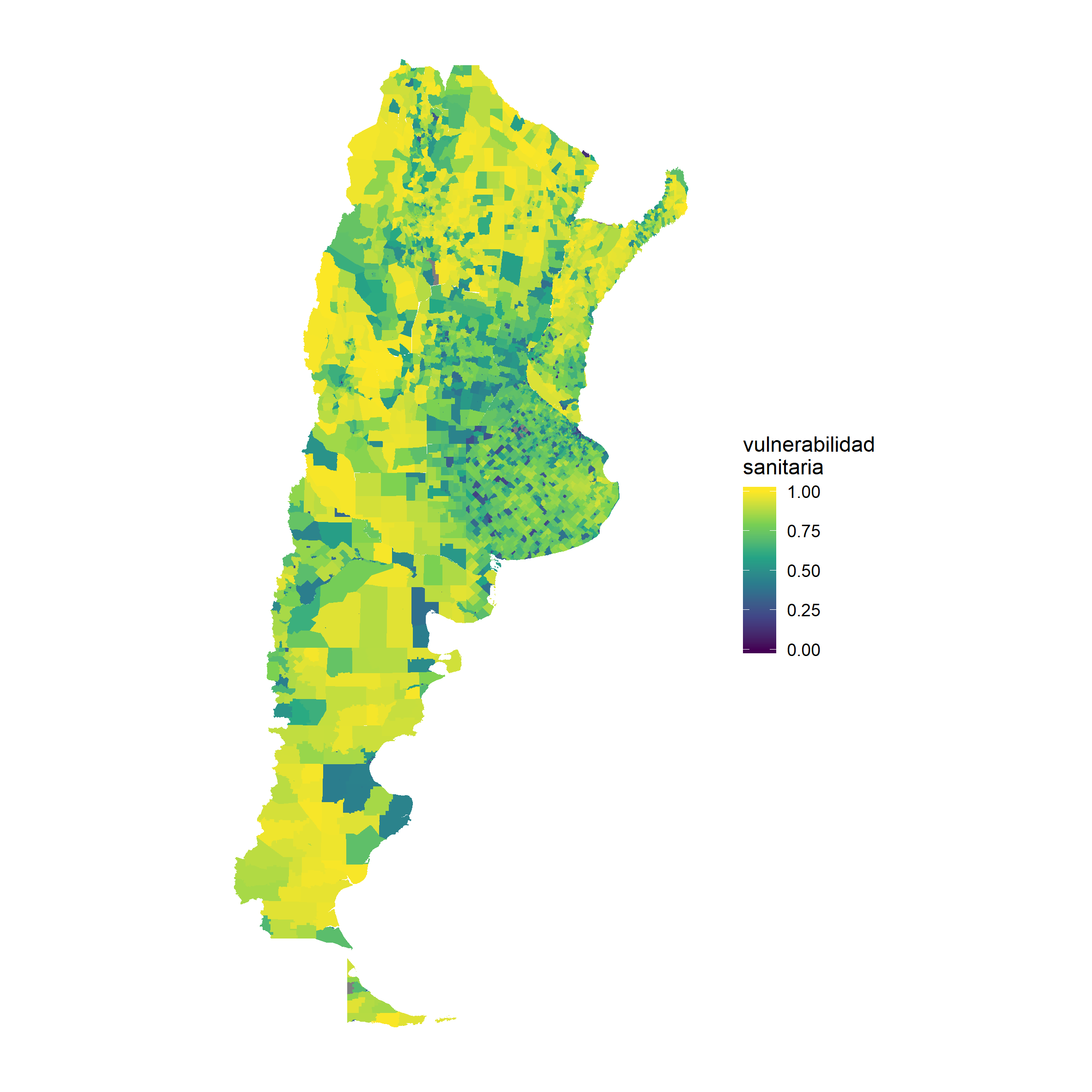} 
	}
	\caption{Índice de Vulnerabilidad Sanitaria a nivel fracción censal (Argentina).}\label{fig:indice_vs_fracc}
\end{figure}

Este mapa arroja una visión más matizada: las regiones detectadas previamente no son tan marcadas. Se observan en la zona central regiones con valores críticos. A su vez, en la zona no central, aparecen ahora sectores mejor rankeados en relación a este indicador.

A su vez, al focalizar la mirada en ciertas áreas (pero mantentiendo el nivel de desagregación por radio), también se logran destacar patrones invisibles en el mapa agregado.  Si se observa la distribución final del Índice de Vulnerabilidad Sanitaria construido para el total del país y para las provincias de Chaco y la CABA se obtiene el siguiente resultado:

\begin{figure}[H]
	{\centering \includegraphics[]{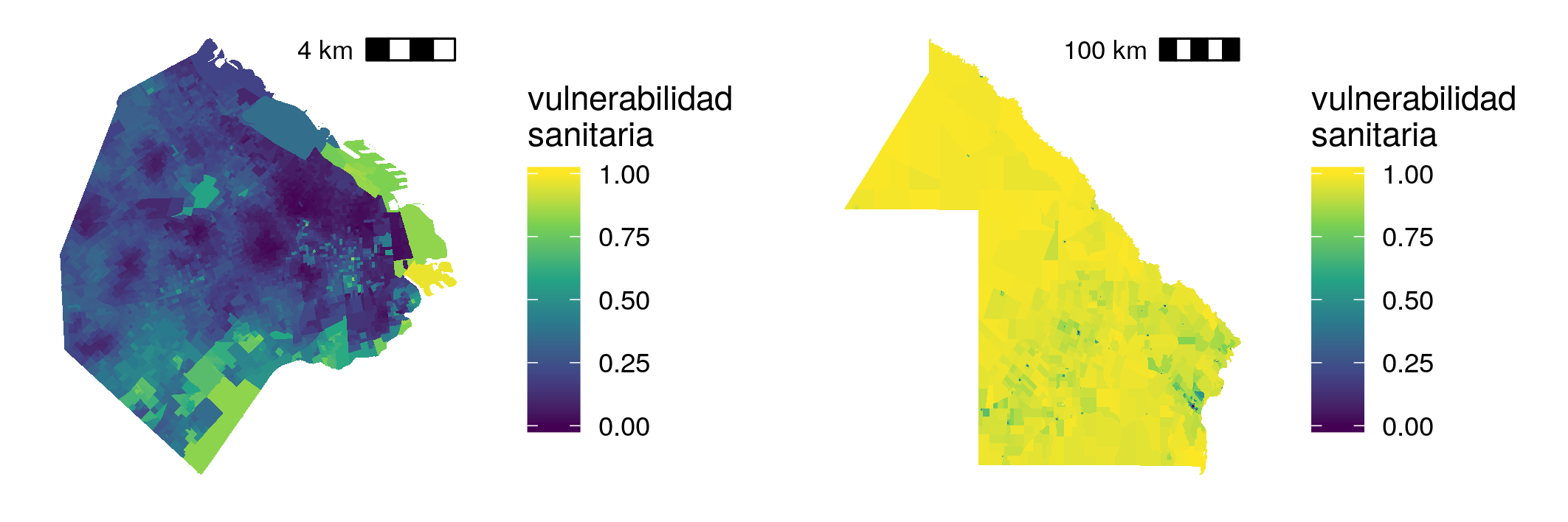} 
		
	}
	\caption{Índice de Vulnerabilidad Sanitaria. CABA (izq.), Chaco (der.)}\label{fig:indice_vs_prov}
\end{figure}

Así, en la CABA es posible detectar nichos de alta vulnerabilidad santiaria y en Chaco existen zonas que presentan valores satisfactorios de este indicador.

\section{Discusión y Conclusión}
\label{conclusion}

La generación de este mapa abre líneas de trabajo que pueden resultar de interés. El primero y más evidente se vincula a la mejora y actualización del Mapa. Una segunda iteración en la limpieza y consolidación de los datos georreferenciados de efectores de salud emerge como una tarea a ser encarada. Al mismo tiempo, la incorporación de nuevas fuentes de datos y dimensiones vinculadas al riesgo y vulerabilidad ambiental resultan de potencial interés para el trabajo futuro.

Un primer resultado tiene que ver con la generación de una metodología de trabajo que resulta replicable y aplicable a otros ámbitos. De esta forma, la construcción de los mapas de distancias a establecimientos de salud podría ser replicada para otros serivicios que son, en principio, potestad del Estado en sus diversos niveles: por ejemplo, educación, seguridad, asistencia social, vivienda, etc. Esto abre a posibilidad de la generación de información con alto nivel de desagregación que sería útil para la toma de decisiones costo-efectivas en la alocación de recursos para la política pública.

A su vez, la Vulnerabilidad Sanitaria puede ser considerada como una dimensión transversal y que afecta la evolución, transmisión y prevalencia de diferentes patologías. De esta forma una segunda línea de trabajo se vincula con la posibilidad de considerar este mapa de Vulnerabilidad Sanitaria (y sus posteriores actualizaciones) como un insumo para el estudio de otras enfermedades infecciosas.


\newpage


\bibliographystyle{abbrv}
\bibliography{biblio}

\newpage

\section{Anexo 1. Descripción de las variables utilizadas en el Índice
	de Nivel Socioeconómico
	(INSE)}
\label{anexo-1.-descripcion-de-las-variables-utilizadas-en-el-indice-de-nivel-socioeconomico-inse}

Se describe en este anexo las variables utilizadas para la construcción
del índice. Para una referencia más exhaustiva, ver~\cite{indec_censo1}. 

\subsection{Relación de parentesco
	(P01):}\label{relacion-de-parentesco-p01}

Indica la relación de cada miembro del hogar con quien ellos hayan
designado como jefe o jefa del mismo. Refiere tanto a las relaciones de
parentesco (sean consanguíneas o no) como a las relaciones de amistad,
de trabajo o de otro tipo. Las categorías son:

\begin{enumerate}
	\def\labelenumi{\arabic{enumi}.}
	\tightlist
	\item
	Jefe(a): persona reconocida como tal por los demás miembros del hogar.
	\item
	Cónyuge: es la pareja de unión legal (unión civil/matrimonio) o la
	pareja de unión ``de hecho'' del jefe o jefa.
	\item
	Hijo(a) / hijastro (a): incluye hijos e hijas biológicos y adoptivos
	legales o de hecho del jefe(a) o criados del jefe o jefa y/o también
	incluye a los hijos e hijas del o la cónyuge aunque no lo sean del
	jefe o jefa.
	\item
	Yerno / nuera: cónyuge o pareja del hijo(a) o hijastro(a) del jefe o
	jefa del hogar.
	\item
	Nieto(a): nietos del jefe o jefa del hogar. Se incluye también a los
	hijos de hijastros(as).
	\item
	Padre / madre / suegro(a): estas relaciones son aplicables también a
	vínculos funcionalmente equivalentes, tales como los de padrastros y
	madrastras con yernos, nietos, etcétera, referidos tanto a los
	matrimonios como a las uniones de hecho.
	\item
	Otros familiares: personas con algún otro tipo de parentesco con el
	jefe o la jefa del hogar (por ejemplo: tíos, etcétera).
	\item
	Otros no familiares: personas que forman parte del hogar y que no
	tienen parentesco con el jefe(a) del hogar (amigos, etcétera) y que no
	están comprendidos en ninguna de las categorías precedentes.
	\item
	Servicio doméstico y sus familiares: persona contratada (legalmente o
	de hecho) para desarrollar tareas relacionadas con el cuidado y
	mantenimiento de la vivienda y/o de los integrantes del hogar. La
	retribución por su trabajo puede ser en dinero o en especie. Incluye a
	sus familiares.
\end{enumerate}

\subsection{Sabe leer y escribir (condición de alfabetismo -
	P07):}\label{sabe-leer-y-escribir-condicion-de-alfabetismo---p07}

Refiere a la capacidad de leer, escribir y comprender una frase sencilla
sobre la vida cotidiana en cualquier idioma. Se requiere el conocimiento
de ambas capacidades.

\begin{enumerate}
	\def\labelenumi{\arabic{enumi}.}
	\tightlist
	\item
	Si
	\item
	No
\end{enumerate}

\subsection{Asiste o asistió a un establecimiento educativo
	(P08):}\label{asiste-o-asistio-a-un-establecimiento-educativo-p08}

Concurrencia actual o pasada (si asiste/asistió) a un establecimiento
reconocido del sistema de la enseñanza formal. Comprende a los
establecimientos del sector estatal o privado. Según el Ministerio de
Educación es importante destacar que para el cálculo de indicadores
educativos y asistencia escolar, las edades deberían ser recalculadas al
30 de junio. Por ese motivo, actualmente se está trabajando para
incorporar una variable ``edad calculada al 30 de junio'', con la que se
podrán obtener las tasas e indicadores correspondientes.

\begin{enumerate}
	\def\labelenumi{\arabic{enumi}.}
	\tightlist
	\item
	Asiste
	\item
	Asistió
	\item
	Nunca asistió
\end{enumerate}

\subsection{Nivel educativo que cursa o cursó
	(P09):}\label{nivel-educativo-que-cursa-o-curso-p09}

Nivel que el censado cursó o está cursando en Argentina o en el exterior
en la fecha del Censo. Las categorías para la variable Nivel educativo
son:

\begin{enumerate}
	\def\labelenumi{\arabic{enumi}.}
	\tightlist
	\item
	Nivel inicial (jardín/preescolar): nivel de la estructura implementada
	a través de la Ley Federal de Educación y la Ley de Educación Nacional
	que comprende a los/as niños/as desde los cuarenta y cinco (45) días
	hasta los cinco (5) años de edad inclusive, siendo obligatorio este
	último año.
	\item
	Primario: comprende los niveles de escolaridad primaria, de carácter
	obligatorio, cuya duración puede ser de seis o siete años (1° a 7°
	grado o 1° a 6° grado).
	\item
	EGB: nivel de la estructura implementada a través de la Ley Federal de
	Educación, actualmente vigente en la provincia de Buenos Aires y otras
	provincias del país. La duración es de nueve años y se encuentra
	compuesta por tres ciclos de tres años cada uno, EGB 1 (1° a 3°
	grado); EGB 2 (4° a 6° grado) y EGB 3 (7° a 9° grado).
	\item
	Secundario: nivel de escolaridad media o secundaria aún vigente. La
	duración puede ser de cinco o seis años (1° a 5° año o 1° a 6° año).
	Las escuelas técnicas/industriales y las dependientes de la
	Universidad, también tienen una duración de seis años; este sistema
	aún está vigente en la Ciudad Autónoma de Buenos Aires y en algunas
	provincias de nuestro país.
	\item
	Polimodal: nivel de la estructura implementada a través de la Ley
	Federal de Educación, actualmente vigente en la provincia de Buenos
	Aires y otras provincias del país. La duración es de tres años.
	\item
	Superior no universitario: nivel de estudios que se realiza en
	instituciones de educación terciaria no universitaria, estatales o
	privados, con planes de estudios aprobados por el Ministerio de
	Educación (de la Nación o de las provincias). Incluye los profesorados
	de nivel inicial, de adultos, de enseñanza especial (sordomudos,
	ciegos, sordos, etcétera) y educación física, historia, letras,
	etcétera. Comprende también especialidades no docentes, por ejemplo:
	especialización técnica industrial, periodismo, turismo, computación,
	bellas artes y la formación de oficiales de las fuerzas armadas.
	\item
	Universitario: nivel de estudios que se realiza en universidades
	nacionales, provinciales o privadas. Comprende exclusivamente las
	carreras que otorgan títulos profesionales (abogado, agrimensor,
	fonoaudiólogo, ingeniero, profesor, etcétera) y las licenciaturas (en
	ciencias de la educación, letras, matemática, sistemas, etcétera).
	\item
	Post-universitario (especialización, maestría o doctorado): nivel de
	estudios que comprende a las carreras de especialización, maestrías y
	doctorados acreditadas por la Comisión Nacional de Evaluación y
	Acreditación Universitaria (CONEAU) o por entidades privadas
	debidamente reconocidas por el Ministerio de Educación. Para acceder a
	este tipo de estudio se requiere contar con título universitario de
	grado. Este tipo de educación se desarrolla en instituciones
	universitarias y en centros de investigación e instituciones de
	formación profesional superior que suscribieron convenios con
	universidades a esos efectos. La realización de carreras de Postgrado
	conduce al otorgamiento del título académico de Especialista, Magíster
	o Doctor.
	\item
	Educación especial (para personas con discapacidad): modalidad del
	Sistema Educativo Nacional destinada a asegurar el derecho a la
	educación de las personas con discapacidades, temporales o
	permanentes, en todos los niveles y modalidades del Sistema Educativo.
	La educación especial brinda atención educativa en todas aquellas
	problemáticas específicas que no puedan ser abordadas por la educación
	común. Son escuelas donde pueden concurrir niños con discapacidad
	mental (leve, moderada o severa), con discapacidad sensorial, con
	discapacidad motora sin compromiso intelectual. Es el conjunto de
	servicios, técnicas, estrategias, conocimientos y recursos pedagógicos
	- dentro del Sistema Educativo Nacional- orientados a la atención de
	las personas con necesidades educativas especiales a causa de una
	discapacidad.
\end{enumerate}

\subsection{Completó ese nivel (P10):}\label{completo-ese-nivel-p10}

Refiere a la aprobación del último año de estudio y a la obtención del
diploma o certificado correspondiente a un determinado nivel.

\begin{enumerate}
	\def\labelenumi{\arabic{enumi}.}
	\tightlist
	\item
	Sí
	\item
	No
	\item
	Ignorado
\end{enumerate}

\subsection{Nivel educativo:}\label{nivel-educativo}

A partir de estas tres variables (P08, P08, P10) se construye la
variable final que tiene las siguientes categorías:

\begin{enumerate}
	\def\labelenumi{\arabic{enumi}.}
	\tightlist
	\item
	Primario incompleto
	\item
	Primario completo
	\item
	Secundario incompleto
	\item
	Secundario completo
	\item
	Universitario incompleto
	\item
	Universitario completo o superior
\end{enumerate}

\subsection{Condición de actividad
	(CONDACT)}\label{condicion-de-actividad-condact}

Comprende a la población de 14 o más años que, en el período de
referencia adoptado por el Censo, estuvo:

\begin{enumerate}
	\def\labelenumi{\arabic{enumi}.}
	\tightlist
	\item
	Ocupada: población que durante por lo menos una hora en la semana
	anterior a la fecha de referencia del censo desarrolló cualquier
	actividad (paga o no) que genera bienes o servicios para el
	``mercado''. Incluye a quienes realizaron tareas regulares de ayuda en
	la actividad de un familiar, reciban o no una remuneración por ello y
	a quienes se hallaron en uso de licencia por cualquier motivo. Se
	excluye de la actividad económica los trabajos voluntarios o
	comunitarios que no son retribuidos de ninguna manera.
	\item
	Desocupada: es la población que no hallándose en ninguna de las
	situaciones descriptas, desarrolló, durante las cuatro semanas
	anteriores al día del censo, acciones tendientes a establecer una
	relación laboral o iniciar una actividad empresaria (tales como
	responder o publicar avisos en los diarios u otros medios solicitando
	empleo, registrarse en bolsas de trabajo, buscar recursos financieros
	o materiales para establecer una empresa, solicitar permisos y
	licencias para iniciar una actividad laboral, etcétera).
	\item
	Económicamente inactiva: comprende a la población de 14 y más años no
	incluida en la población económicamente activa. Incluye a jubilados,
	estudiantes y otras situaciones.
\end{enumerate}

\subsection{Hacinamiento (INDHAC):}\label{hacinamiento-indhac}

Representa el cociente entre la cantidad total de personas del hogar y
la cantidad total de habitaciones o piezas de que dispone el mismo (sin
contar baño/s y cocina/s). Las categorías son:

\begin{enumerate}
	\def\labelenumi{\arabic{enumi}.}
	\tightlist
	\item
	Hasta 0,50 personas por cuarto
	\item
	De 0,51 a 1,00 personas por cuarto
	\item
	De 1,01 a 1,50 personas por cuarto
	\item
	De 1,51 a 2,00 personas por cuarto
	\item
	De 2,01 a 3,00 personas por cuarto
	\item
	Más de 3,00 personas por cuarto
\end{enumerate}

\subsection{Tenencia de la vivienda y propiedad del terreno
	(PROP):}\label{tenencia-de-la-vivienda-y-propiedad-del-terreno-prop}

Refiere al conjunto de normas jurídico-legales o de hecho en virtud de
los cuales el hogar ocupa toda o parte de una vivienda. Las categorías
son:

\begin{enumerate}
	\def\labelenumi{\arabic{enumi}.}
	\tightlist
	\item
	Propietario de la vivienda y del terreno: la vivienda y el terreno en
	el que está ubicada la misma, pertenecen a alguno/s de los integrantes
	del hogar. El hogar tiene capacidad (garantizada legalmente) para
	disponer de la vivienda y del terreno, aún cuando alguno de ellos esté
	pendiente de pago o tenga posesión de los mismos sin haber
	escriturado. El propietario de una vivienda tipo ``departamento'',
	también lo es del terreno.
	\item
	Propietario sólo de la vivienda: la vivienda (pero no el terreno en el
	que está ubicada) pertenece a alguno/s de los integrantes del hogar.
	El hogar tiene capacidad (garantizada legalmente) para disponer de la
	vivienda aún cuando ésta esté pendiente de pago o tenga posesión de la
	misma sin haber escriturado.
	\item
	Inquilino: el hogar paga, por la utilización de toda o parte de una
	vivienda, una cantidad en dinero o en especie (anual, mensual,
	quincenal, etcétera), independientemente de que medie un contrato
	legal.\\
	\item
	Ocupante por préstamo: el hogar utiliza la vivienda que le es
	facilitada gratuitamente por el propietario. La vivienda no es
	propiedad de ninguno de los ocupantes, no está en régimen de alquiler
	y no existe contraprestación alguna por el uso de la misma.
	\item
	Ocupante por trabajo: el hogar utiliza la vivienda que es facilitada
	gratuita o semi-gratuitamente por el patrón, organismo u empresa donde
	trabaja alguno de los miembros del hogar en virtud de su relación
	laboral. Un ejemplo son los porteros, serenos, caseros, trabajadores
	rurales, etcétera.
	\item
	Otra situación: el hogar utiliza la vivienda con una modalidad que no
	se ajusta a ninguna de las anteriores
\end{enumerate}

\subsection{Al menos un indicador NBI
	(ALGUNBI)}\label{al-menos-un-indicador-nbi-algunbi}

\begin{enumerate}
	\def\labelenumi{\arabic{enumi}.}
	\setcounter{enumi}{-1}
	\tightlist
	\item
	No
	\item
	Sí
\end{enumerate}

\subsection{Calidad de los materiales
	(INMAT):}\label{calidad-de-los-materiales-inmat}

Refiere a la calidad de los materiales con que están construidas las
viviendas (material predominante de los pisos y techos), teniendo en
cuenta la solidez, resistencia y capacidad de aislamiento, así como
también su terminación. Se clasifica la calidad de los materiales en:

\begin{enumerate}
	\def\labelenumi{\arabic{enumi}.}
	\tightlist
	\item
	Calidad I: la vivienda presenta materiales resistentes y sólidos tanto
	en el piso como en techo; presenta cielorraso.
	\item
	Calidad II: la vivienda presenta materiales resistentes y sólidos
	tanto en el piso como en el techo. Y techos sin cielorraso o bien
	materiales de menor calidad en pisos
	\item
	Calidad III: la vivienda presenta materiales poco resistentes y
	sólidos en techo y en pisos.
	\item
	Calidad IV: la vivienda presenta materiales de baja calidad en pisos y
	techos
\end{enumerate}

\subsection{Calidad de conexión a servicios básicos
	(INCALSERV):}\label{calidad-de-conexion-a-servicios-basicos-incalserv}

Refiere al tipo de instalaciones con que cuentan las viviendas para su
saneamiento. Para este indicador, se utilizan las variables procedencia
del agua y tipo de desagüe. Las categorías clasificatorias son

\begin{enumerate}
	\def\labelenumi{\arabic{enumi}.}
	\tightlist
	\item
	Calidad satisfactoria: refiere a las viviendas que disponen de agua a
	red pública y desagüe cloacal.
	\item
	Calidad básica: describe la situación de aquellas viviendas que
	disponen de agua de red pública y el desagüe a pozo con cámara
	séptica.
	\item
	Calidad insuficiente: engloba a las viviendas que no cumplen ninguna
	de las 2 condiciones anteriores.
\end{enumerate}

\subsection{Calidad constructiva de la vivienda
	(INCALCONS):}\label{calidad-constructiva-de-la-vivienda-incalcons}

Este indicador se construye a partir de la calidad de los materiales con
los que está construida la vivienda y las instalaciones internas a
servicios básicos (agua de red y desagüe) de las que dispone.

\begin{enumerate}
	\def\labelenumi{\arabic{enumi}.}
	\tightlist
	\item
	Calidad satisfactoria: refiere a las viviendas que disponen de
	materiales resistentes, sólidos y con la aislación adecuada. A su vez
	también disponen de cañerías dentro de la vivienda y de inodoro con
	descarga de agua.
	\item
	Calidad básica: no cuentan con elementos adecuados de aislación o
	tienen techo de chapa o fibrocemento. Al igual que el anterior,
	cuentan con cañerías dentro de la vivienda y de inodoro con descarga
	de agua.
	\item
	Calidad insuficiente: engloba a las viviendas que no cumplen ninguna
	de las 2 condiciones anteriores.
\end{enumerate}
\end{document}